\documentclass[10pt,twocolumn,aps,amssymb,amsmath,a4paper,floatfix,showpacs]{revtex4}

\newcommand{\BE}{\begin{equation}}
\newcommand{\EE}{\end{equation}}

\def\bq{\begin{equation}}
\def\eq{\end{equation}}

\usepackage{color,graphicx,ulem}
\usepackage{bm}

\begin{document}

\title{Finite Larmor radius effects on non-diffusive tracer transport in a zonal flow}

\author{K. Gustafson}
\email{kgustaf@umd.edu}
\affiliation{Department of Physics, University of Maryland, College Park, Maryland 20742-3511}

\author{D. del-Castillo-Negrete}
\affiliation{Fusion Energy Division, Oak Ridge National Laboratory, Oak Ridge, TN 37831}

\author{W. Dorland}
\affiliation{Department of Physics, University of Maryland, College Park, MD 20742-3511}

\date{\today}

\begin{abstract}

Finite Larmor radius (FLR) effects on non-diffusive transport in a prototypical zonal flow with drift waves 
are studied in the context of a simplified chaotic transport model.  The model consists of a superposition 
of drift waves of the linearized Hasegawa-Mima equation and a zonal shear flow perpendicular to the 
density gradient.  High frequency FLR effects are incorporated by gyroaveraging the $\bf{E}\times\bf{B}
$ velocity.  Transport in the direction of the density gradient is negligible and we therefore focus on 
transport parallel to the zonal flows. A prescribed asymmetry produces strongly asymmetric non-
Gaussian PDFs of particle displacements, with L\'evy flights in one direction but not the other.  For $k_
\perp \rho_{th}=0$, where $k_\perp$ is the characteristic wavelength of the flow and $\rho_{th}$ is the 
thermal Larmor radius, a transition is observed in the scaling of the second moment of particle 
displacements, $\sigma^2 \sim t^\gamma$.  The transition separates ballistic motion, $\gamma \approx 2
$, at intermediate times from super-diffusion, $\gamma =1.6$, at larger times. This change of scaling is 
accompanied by the transition of the PDF of particle displacements from algebraic decay to exponential 
decay.  However, FLR effects seem to eliminate this transition. In all cases, the Lagrangian velocity 
autocorrelation function exhibits non-diffusive algebraic decay, ${\cal C} \sim \tau^{-\zeta}$, with $
\zeta=2-\gamma$ to a good approximation. The PDFs of trapping and flight  events show clear evidence 
of algebraic scaling with decay exponents depending on the value of $k_{\perp}\rho_{th}$. The shape 
and spatio-temporal self-similar anomalous scaling of the PDFs of particle displacements are 
reproduced accurately with a neutral, $\alpha=\beta$, asymmetric effective fractional diffusion model 
where $\alpha$ and $\beta$ are the orders of the spatial and temporal fractional derivatives.

\end{abstract}



\maketitle

\section{Introduction}
\label{sec:intro}

Plasma turbulence presents a challenge to multiscale models of transport in applications such as 
magnetic fusion confinement, stellar accretion disks and galactic dynamos.  Simulations of turbulent 
transport involve nonlinear interactions at disparate scales, which often makes numerical computations 
expensive and analytic methods intractable.  As an alternative, one may consider models of 
intermediate complexity that incorporate important aspects of transport within a relatively simple reduced 
description. In this paper we follow this approach and present a numerical study of the role of  finite 
Larmor radius (FLR) effects on non-diffusive poloidal transport in zonal shear flows using a reduced $
{\bf E} \times {\bf B}$ Hamiltonian test particle transport model.

Following Ref.~\cite{dCN00}, we model the flow as a superposition of a shear flow and drift waves 
obtained from the linearized Hasegawa-Mima (HM) equation~\cite{Hasegawa78}.  Test particle 
characteristics in this flow are generally not integrable and exhibit chaotic advection, also known as 
Lagrangian turbulence, which reproduces key ingredients of particle transport in more complex flows. 
High frequency FLR effects are incorporated by solving the test particle equations of motion for the 
gyroaveraged ${\bf E} \times {\bf B}$ velocity. As demonstrated by Ref.~\cite{Lee87}, we compute the 
gyroaverage using a discrete $N$-polygon approximation.

We adopt a statistical approach and apply non-diffusive transport diagnostics to large ensembles of 
particles. One of the simplest diagnostics is the scaling of the second moment of particle displacements, 
$\sigma^2(t) = \langle [ \delta y - \langle \delta y \rangle]^2 \rangle$, where $\delta y =\delta y(t)$ denotes 
the particle's displacement and $\langle \, \rangle$ denotes the ensemble average. In the standard 
diffusion case, $\sigma^2(t) \sim t$, linear scaling allows the definition of an effective diffusivity as the 
ratio $D_{eff}=\sigma^2(t)/(2 t)$ in the limit of large $t$. However, in the case of non-diffusive transport, $
\sigma^2(t) \sim t^\gamma$ with $\gamma \neq 1$. When $0<\gamma<1$, the growth of the variance is 
slower than diffusion and transport is sub-diffusive.   When $1<\gamma<2$ transport is super-
diffusive, which means the spreading is faster than diffusion, and the displacements may be L\'evy 
flights~\cite{Metzler00}. In both super- and sub-diffusion, characterization of transport as a diffusive 
process with an``effective diffusivity" $D_{eff}$ breaks down because $D_{eff} \rightarrow 0$ when $0<
\gamma<1$, and  $D_{eff} \rightarrow \infty$ when $1<\gamma<2$.  Other measures of non-diffusive 
transport, which will be discussed in detail later, include non-Gaussianity of the probability distribution of  
displacements (propagator), slow decay of the Lagrangian velocity autocorrelation function, the 
presence of long jumps (L\'evy flights) and long waiting times, and the non-local ({\it i.e.}, non-Fickian) 
dependence of fluxes on gradients.  A general review of non-diffusive transport can be found in Ref.~
\cite{Bouchard90}, and discussions focusing on plasmas can be found in Ref.~\cite
{Balescu05,dCN08a}.

\begin{figure}
\begin{center}
\includegraphics[width=3in]{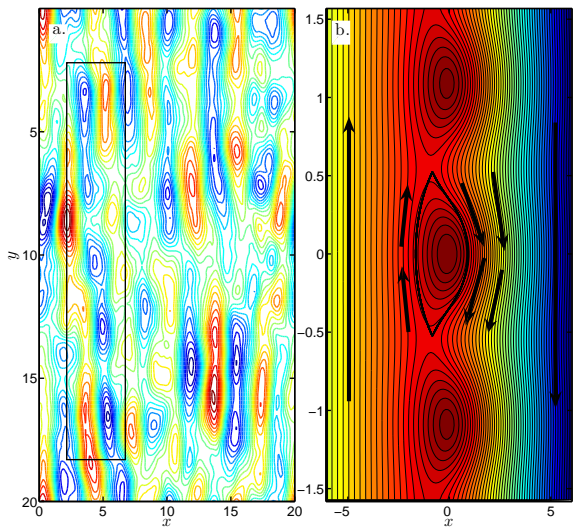}
\end{center}
\caption{(Color online) Contour plots of electrostatic potential $\phi$. Panel (a) shows a snapshot of the potential 
obtained from a direct numerical simulation of the Hasegawa-Mima equation~(\ref{eq:analogy-1}).
Panel (b) shows $\phi$ at a fixed time according to the chaotic Hamiltonian transport model in
Eq.~(\ref{eq:stat-6}).  The thick line limiting the central vortex in (b) is the separatrix. Particles inside the 
separatrix are trapped, and, as the arrows show, particles outside the separatrix are transported by the 
zonal flow. The Hamiltonian model in (b) provides a reduced description  of ${\bf E} \times {\bf B}$
transport dominated by vortices and zonal flows as highlighted by the rectangle in (a).}
\label{fig:NLcomparison}
\end{figure}

Test particle transport in HM flows, as in Fig.~\ref{fig:NLcomparison}(a), has been 
studied in Refs.~\cite{Horton81,Manfredi96,Manfredi97,BGZ97,Annibaldi00,dCN00,Annibaldi02}. In 
Ref.~\cite{dCN00}, which did not include FLR effects, it was shown that zonal flows give rise to L\'evy 
flights and strongly asymmetric non-Gaussian PDFs of particle displacements.  
References~\cite{Manfredi96,Manfredi97} addressed the  role of FLR effects but restricted attention to 
diffusive transport. More recently, Ref.~\cite{Annibaldi02} considered FLR effects in non-diffusive 
transport in HM turbulence and concluded that the exponent $\gamma$ does not change appreciably 
with the Larmor radius  but that the effective diffusion coefficient is reduced.  There is a very close 
connection between drift waves as described by the HM equation and Rossby waves as described by 
the quasigeostrophic equation, see for example Ref.~\cite{Horton94}.  Therefore, statistical test particle 
studies in fluid mechanics, such as Refs.~\cite{dCN98,Kovalyov00}, are in principle applicable to drift 
wave transport.

The main new results presented here, which to our knowledge have not been reported in the 
literature before, include: (i) a transition from algebraic to exponential decay in the tails of PDFs of 
particle displacements accompanied by a transition from ballistic ($\gamma \approx 2$) to super-
diffusive ($1<\gamma < 2$)  
transport; (ii) a numerical study of the role of FLR on the Lagrangian velocity autocorrelation function 
and on the particle trapping and particle flight PDFs; (iii) the construction of a effective
fractional diffusion model that reproduces the shape and the spatio-temporal anomalous self-similar 
scaling of the PDF of particle displacements.  In recent years, fractional diffusion models have been 
applied to describe non-diffusive plasma transport, {\it e.g.} Refs.~\cite
{Chechkin02,dCN04,dCN06,Garcia06,dCN08b,Calvo08}. Although the present work focuses on a 
prototypical model of transport, the diagnostics used and the non-diffusive phenomenology discussed 
here might be of relevance to the study of transport in more general flows dominated by coherent 
structures like zonal flows and eddies.  Despite the fact that these coherent structures are ubiquitous in 
simulations and 
experiments~\cite{Horton94,Jenko02,Furno08}, their influence on non-diffusive transport is not well 
understood. In this regard, Ref.~\cite{Hauff07b} showed evidence of non-diffusive transport in 
gyrokinetic turbulence for ``intermediate" simulation times.

The rest of the paper is organized as follows. In Sec.~\ref{sec:model} the ${\bf E} \times {\bf 
B}$ transport model with and without FLR effects is explained.  Section \ref{sec:method} shows a 
benchmark of the numerical method against an exact  solution for the particle propagator in a parallel 
flow. Section~\ref{sec:diagnostics} presents a summary of Lagrangian diagnostics to study non-diffusive 
transport.  The main numerical results are presented in Sec.~\ref{sec:results}. Section~\ref
{sec:modeling} describes the anomalous self-similarity properties of the PDF of particle displacements 
and presents an effective fractional diffusion model. Section \ref{sec:conclusion} contains the 
conclusions.

\section{Transport model}
\label{sec:model}

We follow a Lagrangian approach to study transport and consider large ensembles of discrete particles
moving in a prescribed flow. We limit attention to test particles, neglecting
self-consistency effects and assuming that the particles are transported by the flow without modifying it.
When finite Larmor radius (FLR) effects can also be neglected, the
dynamics are determined by a drift equation which, in the ${\bf E}\times {\bf B}$
approximation, is
\bq
\label{eq:ExB}
\frac{d {\bf r}}{dt} = \frac{{\bf E}\times {\bf B}}{B^2} \, ,
\eq
where ${\bf r}=(x,y)$ denotes the particle position, ${\bf E}$ is
the electrostatic field, and ${\bf B}$ is the magnetic field. Writing ${\bf B}=B_0  \hat{\bf z}$, and
${\bf E}=-\nabla \phi(x,y,t)$, Eq.~(\ref{eq:ExB}) can be equivalently written as the Hamiltonian
dynamical system
\bq
\label{eq:hamilton}
\frac{d x}{d t} = - \frac{\partial \phi}{\partial
y}\, ,\qquad
\frac{d y}{d t} =  \frac{\partial  \phi}{\partial x}\, ,
\eq
where the electrostatic potential is analogous to the Hamiltonian, and the spatial coordinates are the
canonical conjugate phase space variables.

For relatively high energy particles or for a flow varying relatively rapidly in space, the zero Larmor 
radius approximation fails and it is necessary to incorporate
FLR effects.   A simple, natural way of doing this is to substitute the
${\bf E}\times {\bf B}$ flow on the right hand side of Eq.~(\ref{eq:hamilton}), which is evaluated at the
location of the guiding center, by its value
averaged over a ring of radius $\rho$, where  $\rho$ is the Larmor radius \cite{Lee87}.
Formally, the procedure is given by
\bq
\label{eq:gavhamilton}
\frac{d x}{dt} = -\left \langle \frac{\partial \phi}{\partial y}  \right \rangle_\theta\, ,
\qquad
\frac{d y}{dt} = \left \langle \frac{\partial \phi}{\partial x} \right \rangle_\theta \,
\eq
where the gyroaverage, $\left \langle \,  \right \rangle_\theta$, is defined as
\bq
\label{eq:gyroaverage}
\langle  \Psi \rangle_\theta \equiv \frac{1}{2 \pi} \int_0^{2 \pi} \Psi \left(x + \rho \cos
\theta,  y + \rho \sin \theta \right) d \theta \, .
\eq
This is a good approximation provided the gyrofrequency is greater than other frequencies in the 
system.  

In the HM model for drift waves the electrostatic potential is determined
from \cite{Hasegawa78}
\bq
\label{eq:analogy-1}
\left[\partial_t +
\left({\bf z} \times \nabla \phi\right) \cdot \nabla \right ]\, \left( \nabla^2 \phi - \phi - \beta  x \right )
 =0\, ,
\eq
where the $x$ coordinate corresponds to the direction of the density
gradient driving the drift-wave instability, and $y$ corresponds to the direction of propagation of the
drift-waves. In
toroidal geometry, $x$ is analogous to a normalized coordinate along the minor radius, and $y$ is a
poloidal-like coordinate. Here we assume a slab approximation and treat $(x,y)$ as Cartesian 
coordinates. The parameter  $\beta=n_0(x) ' / n_0(x)$ measures the scale length of the density gradient.
We model the electrostatic potential (test particle Hamiltonian) as a superposition of an equilibrium 
zonal shear flow, $\varphi_0 (x)$, and the corresponding eigenmodes of Eq.~(\ref{eq:analogy-1}), $
\varphi_j(x)$, with perpendicular wave numbers, $k_{\perp j}$, and frequencies,
$c_j k_{\perp j}$,
\bq
\label{eq:model-2}
\phi= \varphi_0(x)
+ \sum_{j=1}^{N} \varepsilon_j \,
\varphi_j(x) \, \cos k_{\perp j}(y - c_j t) \ .
\eq

We consider a monotonic zonal flow of the form
\bq
\label{eq:model-3}
v_{y,0}(x) = \tanh (x) \, .
\eq
In this case, depending on the parameter values, there is a band of
unstable modes bounded by two regular neutral modes  with eigenfunctions \cite{dCN00}
\bq
\label{eq:stat-3}
\begin{array}{cccc}
\varphi_{j}=\left[1+\tanh x \right] ^{\frac{1-c_{j}}{2}}
\left[1-\tanh x \right] ^{\frac{1+c_{j}}{2}}  \ .
\end{array}
\eq
Since these modes are neutral, $c_1$ and $c_2$ are real and the corresponding values of $k_{\perp j}$ 
are
obtained from the linear dispersion relation.
Neutral modes are important because they describe dynamics near marginal stability.
Following Ref.~\cite{dCN00}, we consider a traveling wave
perturbation of the first neutral mode.  The electrostatic potential in the co-moving reference
frame of the neutral mode takes the form
\begin{eqnarray}
\label{eq:stat-6}
\phi= \ln \left( \cosh x \right)  +
\varphi_1(x) \, [ \varepsilon_1\, \cos k_{\perp 1} y  + \nonumber\\ 
\varepsilon_2\, \cos (k_{\perp 2} y  - \omega t)]  - c_1 x
 .
\end{eqnarray}
The first term on the right hand side of Eq.~(\ref{eq:stat-6}) is the potential of the shear flow in
Eq.~(\ref{eq:model-3}), and $\omega$ is the  frequency of the perturbation. 
The wavenumbers perpendicular to the uniform magnetic field, $k_{\perp 1}$ and $k_{\perp 2}$, 
characterize the size of ${\bf E} \times {\bf B}$ eddies, while
$\varepsilon_1$ and $\varepsilon_2$ give the amplitudes of the waves.  When computing $k_{\perp}
\rho_{th}$ to compare the scale length of the eddies in this flow to the thermal gyroradius, we use the 
mean value $k_{\perp}=(k_{\perp 1}+k_{\perp 2})/2$. 

When $\varepsilon_2=0$ the Hamiltonian in Eq.~(\ref{eq:stat-6}) is time independent, and the test
particles follow contours of constant $\phi$ shown in Fig.~\ref{fig:NLcomparison}(b).
In this case, particles inside the separatrix
remain trapped and those outside the separatrix are always untrapped with $\dot{y}>0$ left of the
vortices and $\dot{y}<0$ right of the vortices.  However, when there is a time
dependent perturbation, {\it i.e.} when $\varepsilon_2 \neq 0$ in Eq.~(\ref{eq:stat-6}), the ${\bf E}\times
{\bf B}$ particle trajectories are in general not integrable. In  this case, the separatrix breaks and forms  a
stochastic layer where test particles alternate chaotically  between being untrapped in the zonal flow 
and being trapped inside the vortices. This is the phenomenon of chaotic transport that has been studied 
in both plasmas and fluid systems, see for example 
Refs.~\cite{Horton81,Aref84,Solomon93,dCN98}
and references therein.  As Fig.~\ref{fig:NLcomparison}(a) illustrates, the simple Hamiltonian model in 
Eq.~(\ref{eq:stat-6}) provides a reduced description of ${\bf E} \times {\bf B}$ eddies embedded in a 
background zonal flow in HM turbulence.  

\section{Numerical method}
\label{sec:method}

The zero Larmor radius calculations are based on the Hamitonian-like equations of Eq.~(\ref
{eq:hamilton}). For the numerical integration of these equations we used the second-order
symplectic predictor-corrector scheme of Ref.~\cite{FinndCN01} with a fixed time step of 0.05 and 8
iterations in the predictor-corrector loop.  These parameters were chosen based on numerical
convergence studies and by monitoring the accuracy of energy conservation. For the model parameters
we used $\varepsilon_1 = 0.5$, $\varepsilon_2 = 0.2$, $c_1 = 0.4$, $k_{\perp 1} = 6.0$, $k_{\perp 2} = 
5.0$ and $\omega = 6.0$.   This choice is motivated by Refs.~\cite{dCN98,dCN00} where it was shown 
that, for this set of parameters, test particles exhibit strongly asymmetric, non-Gaussian statistics. As 
such, 
these parameters are a good starting point to study the role of FLR effects on non-diffusive transport.
For the initial conditions we used an ensemble of particles located in the vicinity of the hyperbolic fixed
point of the Hamiltonian at $(x_0,y_0) \sim (-1,-0.5)$. This localization guarantees that a large fraction of 
the particles will stay in the stochastic layer and undergo chaotic transport.  Other choices of initial 
positions can lead to integrable motion with particles permanently either inside the eddies, circling, or 
outside, following the zonal flow.

The only difference between the zero and finite Larmor radius calculations is in the evaluation of the
velocity of the test particle.  Assuming fast gyration in a strong ${\bf B}$ field, the gyroaverage of the ${\bf 
E} \times {\bf B}$ velocity is computed over a circle of radius $\rho$, where $\rho$ is the Larmor radius of 
the particle.
Throughout this paper  we will assume a Maxwellian
equilibrium distribution for the Larmor radii of the test particles of the form
\bq
\label{eq:maxwell}
H(\rho) = \frac{2}{\rho_{th}^2}e^{-\rho^2/\rho_{th}^2} \, ,
\eq
normalized according to $\int_0^\infty H(\rho) \rho d \rho =1$.
For the numerical computation of the gyroaverage we
approximate the circle with an inscribed polygon  with
$N_g$-sides and approximate the integral over the circle as the average over the vertices of the
polygon. This method, widely used in kinetic particle codes ({\it e.g.}~\cite{Lee87}), simply samples the 
field on the gyration arc at a small number
of equally spaced points.  For example, the 8-point (octagon) approximation evaluates the gyroaverage
by considering $N_g=8$ points distributed around the circle in equal increments, {\it i.e.}, at $\theta = \{2 
\pi/8, 2
\pi/7, \ldots 2\pi\}$. If the mean gyroradius, $\langle \rho \rangle = (\sqrt{\pi}/2) \rho_{th}$,
becomes large relative to the typical scale length, $\sim 1/k_{\perp}$,  of the flow, {\it i.e.}, if
$k_{\perp}\rho_{th} \gg 1$, the number of points used to compute the gyroaverage must be increased
to maintain the same level of accuracy.

The error involved in the approximation of the gyroaverage on $N_g$ for a
given value of $k_{\perp}\rho_{th}$ and, therefore, a benchmark for the accuracy of the numerical 
scheme can be studied by considering the following parallel flow in arbitrary geometry
\bq
\label{eq:oned}
\phi= \phi_0 \cos (k_{\perp} x) \, .
\eq
The main object of interest is the probability distribution function of particle displacements, or
propagator, $P=P(y,t | y',t')$, which gives the probability for a particle to be at $y'$
at time $t'$ if it was at $y$ at time $t$.  Since $v_x = 0$ for this choice of $\phi$, we restrict study to the $y
$ direction. The function $P$ depends on $k_{\perp}\rho_{th}$ and
the goal is to study the error in the numerical evaluation of $P$ as function of $k_{\perp}\rho_{th}$ and
the value of $N_g$ used in the approximation of the gyroaverage.
As discussed in Appendix \ref{sec:staticdetails}, the exact propagator for Eq.~(\ref{eq:oned}) is given by
\bq
\label{eq:exact}
P(y,t|y',t') = \frac{1}{U_0 (t-t')}\, {\cal G} (\zeta) \, , \qquad \zeta =
\frac{1}{U_0}\frac{(y-y')}{(t-t')}\, ,
\eq
with
\bq
{\cal G}(\zeta)= \frac{2}{\left(k_{\perp}\rho_{th}\right)^2} \sum_{i=1}^{N_z} \frac{z_i \,
e^{-(z_i/k_{\perp}\rho_{th})^2} }{|J_1 (z_i)|}
\, ,
\eq
where $z_i=z_i(\zeta)$ denotes the  i-th zero of the equation $J_0(z_i)-\zeta=0$. Here, $J_0$ is the 
order zero Bessel function of the first kind. For a given $\zeta$, the number of zeros of this equation is  
$N_z$ which goes to $\infty$ as $\zeta$ goes to zero.  Note also that because the minimum and 
maximum values of $J_0$ are $-0.4025$ and $1$, respectively, no zero exists for $\zeta<-0.4025$ 
or $\zeta>1$.  Therefore, $P$ identically vanishes outside the interval $\zeta \in (-0.4025,1)$.
Despite its apparent complexity, this analytical result provides a valuable benchmark to assess
the accuracy of the gyroaverage computation.

\begin{figure}
\begin{center}
\includegraphics[width=3in]{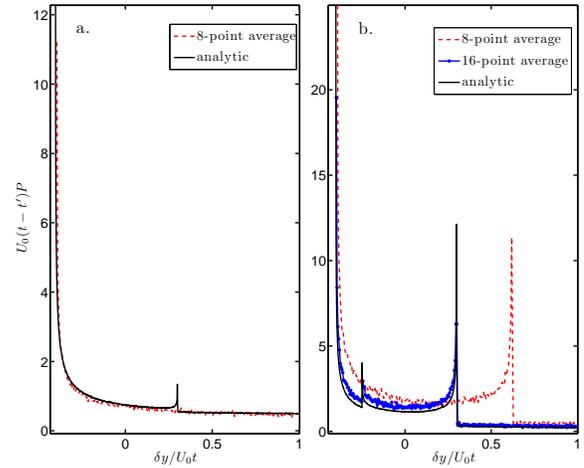}
\end{center}
\caption{(Color online) Particle propagator for finite Larmor radius transport in the parallel shear flow of
Eq.~(\ref{eq:oned}). Panel (a) corresponds to  $k_{\perp}\rho_{th}= 3.0$ and (b) corresponds to
$k_{\perp}\rho_{th}= 5.0$. The solid line denotes the exact analytical  result in Eq.~(\ref{eq:exact}), the 
dashed line and the marked line (shown only in (b)) denote the $8$-point and the $16$-point average 
numerical results, respectively.}
\label{fig:benchmarks}
\end{figure}

Figure~\ref{fig:benchmarks} compares the exact propagator in Eq.~(\ref{eq:exact}) with the  propagator 
obtained from direct numerical integration of the gyroaverage equations of motion in 
Eq.~(\ref{eq:gavhamilton}) for different values of $k_{\perp}\rho_{th}$ and $N_g$. The FLR effects
significantly change the $k_{\perp}\rho_{th}=0$ propagator, which is a $\delta$-function
centered at $\zeta=1$: $P(y,t|y',t')=(1/U_0(t-t')) \delta(\zeta -1)$.
It is observed that for $k_{\perp}\rho_{th} = 3.0$, $N_g=8$ produces relatively good
results, although it misses the small spike in $P$ around $\delta y/ U_0 t \sim 0.25$.
Other $N_g=8$ cases with $k_{\perp}\rho_{th}  \leq 3.0$ (not shown) give nearly exact agreement. 
However, for
$k_{\perp}\rho_{th} = 5.0$, the $N_g=8$ average departs significantly from
the exact result.  This failure means that choosing $N_g>8$, such as $N_g=16$, is necessary.
One is led to conclude that the $N_g=8$ method should not be used for values of
$k_{\perp}\rho_{th} \gtrsim 3.0$.  

This statement is further supported by an assessment of accuracy when
representing $J_0(\iota)$ as a finite sum based on the integral
\bq
J_0(\iota) = \frac{1}{2 \pi} \int_{0}^{2 \pi} \cos (\iota \sin \tau) d\tau \, .
\eq
The Bessel function is used in spectral simulations of the gyrokinetic equation, which gives the spectral 
technique an advantage that we cannot use here.  The Bessel integral representation may be 
discretized and evaluated using different numbers of terms in the sum.  Additional
terms in the sum reduce the error of discretization just as increasing $N_g$ reduces the error of discrete 
gyroaveraging.  When the integral is approximated with  $8$
or $16$  equally spaced points between $0$ and $2\pi$,  the result agrees to $0.1\%$ 
with the value of $J_0(\iota)$
up to $\iota = 3.0$ or $\iota=9.0$, respectively.  For higher values of $\iota$, the approximation
diverges quickly, just as the discrete gyroaverage method diverges from the analytic result for increasing
$k_{\perp}\rho_{th}$.  Based on this, care must be taken in selecting $N_g$ for large
values of $k_{\perp}\rho_{th}$. 
In this paper we restrict attention to  $k_{\perp}\rho_{th} \leq 3.0$ and
use an adaptive $N_g$ technique based on Ref.~\cite{Mishchenko05}.

\section{Diagnostics for non-diffusive transport}
\label{sec:diagnostics}

In this section we review several Lagrangian diagnostics for transport study.
After defining each diagnostic, we recall expected behavior for both diffusive and
non-diffusive transport. These diagnostics have been successfully used in 
transport experiments, models, and simulations in both fluids and plasmas. 
For examples see Refs.~\cite{Solomon93,dCN98,dCN04} and references therein.  To simplify the 
discussion we limit attention to one-dimensional transport, {\it i.e.} transport in the poloidal-like direction 
$y$. 
In the specific transport problem considered in this paper, $y$ is in the direction of the propagation of the 
zonal flow and the drift waves, and is orthogonal to both the density gradient and the magnetic field.  
Generalization of the diagnostics to higher dimensions is straightforward.

\subsection{Statistical moments of particle displacements}

The basic particle data consists of the ensemble $\{ y_i (t)\}$, with $i=1,2, \ldots N_p$, containing the
time evolution of the $y$-coordinate of the $N_p$ test particles in the simulation. From here we define 
the ensemble of particle displacements, $\{ \delta y_i(t) \}$, where $\delta y_i(t)=y_i(t)-y_i(0)$.
The statistical moments of the particle displacements provide one of the simplest and most natural 
characterizations of Lagrangian transport. Of particular interest are the mean $M(t) = \langle \delta y 
\rangle$ and the variance $\sigma^2(t) = \langle \left[ \delta y -  \langle \delta y \rangle \right] ^2 \rangle$ 
where $\langle \,\, \rangle$  denotes ensemble average. In the case of diffusive transport ({\it e.g.}, a 
Brownian random walk), the  moments exhibit asymptotic linear scaling in time, which allows the 
definition of an effective transport  velocity (pinch) $V_{eff}$ and an effective diffusivity $D_{eff}$ 
according to $V_{eff} = \lim_{t \rightarrow \infty} M(t)/t$ and $D_{eff} = \lim_{t \rightarrow \infty} \sigma^2
(t)/2 t$.

However, in the case of nondiffusive transport, the moments display anomalous scaling of the form
\bq
\label{eq:an_sca}
M \sim t^\chi \, , \qquad \sigma^2 \sim t^\gamma \, ,
\eq
with  $\chi \neq 1$ and  $\gamma \neq 1$. If $0<\gamma<1$ the spreading is slower than in
the diffusive case and transport is called sub-diffusive. If $1<\gamma<2$, the spreading is faster
than diffusion and transport is super-diffusive. A similar classification applies for sub-advection ($0<\chi
<1$) and super-advection ($1 < \chi <2$).  In the presence of anomalous scaling, the
introduction of an effective transport velocity or an effective diffusivity is meaningless since these
transport coefficients are either zero (in the sub-advection/sub-diffusion case) or infinite (in the
super-advection/super-diffusion case).  The diagnostics based on the statistical moments are
straightforward to implement.  The key is to look for a scaling region in a log-log plot of the moments as
functions of time, after transients have passed. However, as with the data analyzed below, it is possible 
for the moments to follow different scaling regimes for different time intervals.

\subsection{Particle displacement PDFs: spatial scaling}

The probability distribution function (PDF) of particle displacements, $P(\delta y,t|\delta y', t')$, contains 
all of the statistical information from displacements beyond the first and second moments. By definition, 
$P(\delta y,t|\delta y',t'=t)=\delta(y)$. Numerically, $P$ is constructed from the normalized histogram of 
particle positions at a given time. Formally, $P(\delta y,t|\delta y', t')$ corresponds to the Green's function
determining the distribution of the test particles in terms of the initial particle distribution.
For a Brownian random walk, the central limit theorem implies that $P$ asymptotically
approaches a Gaussian distribution, $P_G$,  that satisfies diffusive scaling, $P_G=t^{-1/2}
G(Y/t^{1/2})$, where $G$ is a Gaussian and $Y=\delta y - \langle \delta y \rangle$. However, a
non-diffusive propagator can exhibit the more general (anomalous) self-similar scaling
\bq
\label{eq:self_sim_sca}
P=t^{-\gamma/2} L(Y/t^{\gamma/2}) \, ,
\eq
where $0<\gamma<2$ and $L$ is a non-Gaussian function. Note that, by construction, the propagator
has zero mean, and the scaling exponent $\gamma$ in
Eq.~(\ref{eq:self_sim_sca}) is the same as the exponent in Eq.~(\ref{eq:an_sca}).
From Eq.~(\ref{eq:self_sim_sca}) it follows that
$P(Y,t) = \lambda^{\gamma/2} P(\lambda^{\gamma/2} Y, \lambda t)$
where $\lambda$ is a real number. Therefore, if the propagator is self-similar,
$P$ is invariant with respect to the space-time renormalization transformation $(Y,t) \rightarrow
(\lambda^{\gamma/2} Y, \lambda t)$, up to a scale factor. 

Equation~(\ref{eq:self_sim_sca}) provides a
useful diagnostic to reveal non-diffusive transport and, in particular, the existence of anomalous self-
similar scaling.  This diagnostic is implemented by plotting the propagator at different times in rescaled
coordinates, {\it i.e.} $t^{\gamma/2} P$ versus $Y/t^{\gamma/2}$.  With self-similar non-diffusive
transport, the plots at different times rescale and collapse into a single function $L$. One of the most
important departures from Gaussianity is algebraic decaying, ``fat'' tails in the
propagator for large
$\delta y$ at fixed $t$,
\bq
\label{eq:zeta}
P \sim \delta y^{-\zeta} \,  .
\eq
When this behavior is found, the value of the scaling exponent $\zeta$ is a useful diagnostic that
characterizes the intermittency of the transport process.

\subsection{Trapping and flight probability distribution functions}

Diffusive transport can be interpreted as a coarse-grained (macroscopic) description of a fine-grained
(microscopic) Brownian random walk. In a similar way, non-diffusive transport can sometimes be viewed 
as
the result of a non-Brownian random walk with a non-Gaussian and/or
non-Markovian \cite{SZK93} underlying stochastic process.  Trapping and flight probability distribution
functions are two useful diagnostics for the characterization of non-Brownian random walks. Given a
particle trajectory, $y_i(t)$, a trapping event is defined a portion of the trajectory during which the
particle stays on an eddy. Flight events are portions that are not trapping events. Thus, each particle
orbit in the ensemble of initial conditions may be decomposed as a sequence of trapping and flight 
events.

Numerically, the events are detected by tracking reversals in the Lagrangian acceleration of particles.
From the histograms of trapping and flight events one may construct the probability distribution functions
of trapping events, $\psi(t)$, and flight events, $\lambda(y)$. Indications of non-diffusive transport can
be explored by studying the  departures of $\lambda(y)$ and $\psi(t)$ from the Gaussian and
exponential dependencies characteristic of Brownian random walks. Of particular interest is the 
presence
of asymptotic algebraic scaling of the form,
\bq
\label{eq:trapflight}
\psi \sim t^{-\nu}\, , \qquad \lambda \sim y^{-\mu} \, .
\eq
When $\mu <1$ the mean waiting time, $\int t \psi dt$,  is infinite and no characteristic
temporal scale exists.  In the L\'evy flight regime $\mu<3$, and therefore the second moment, 
$\int y^2\lambda dy$, diverges and no characteristic spatial scale exists.  The PDFs of flight  and 
trapping events are in principle interesting because of their connection to the continuous time  random 
walk (CTRW) model, which, in the fluid continuum limit, can be described using fractional diffusion 
equations \cite{MontrollWeiss65,Montroll84,Metzler00}. 

\subsection{Lagrangian velocity autocorrelation function}

Further insights into non-diffusive transport can be gained by looking at the Lagrangian velocity
autocorrelation function
$C(\tau) = \langle v_y(\tau) v_y(0) \rangle$ where $v_y$ is the Lagrangian velocity of a particle.
The Green-Kubo relation, $d\sigma^2/dt = 2\int_0^t C(\tau)d\tau$,
relates the velocity autocorrelation function to the variance of displacements. When $C$ decays fast
enough so that the integral
converges, this relation can be used to define an effective diffusivity  according to
$D_{eff}=\int_0^\infty C(\tau) d \tau$. However,
when $C$ has algebraic decay of the form
\bq
\label{eq:kappa}
C(\tau) \sim \tau^{-\kappa} \, ,
\eq
with $\kappa <1$, the integral diverges and the concept of effective diffusivity loses meaning.
For super-diffusive transport, $\sigma^2 \sim t^\gamma$ implies $\gamma=2-\kappa$.

\section{Numerical results}
\label{sec:results}

For the Lagrangian statistics we consider ensembles of $N=8\times10^4$ test particles, and integrate
the equations of motion, with and without FLR effects, up to $t=5.2\times10^3$.
The zero Larmor radius results were obtained from the numerical integration of the guiding
center equations in Eq.~(\ref{eq:gavhamilton}) with the Hamiltonian in Eq.~(\ref{eq:stat-6}) with
$\varepsilon_1 = 0.5$, $\varepsilon_2 = 0.2$, $c_1 = 0.4$, $k_{\perp 1} = 6.0$, $k_{\perp 2} = 5.0$, $
\omega = 6.0$. The same Hamiltonian and parameter values were used in the FLR ($0<k_{\perp}\rho_
{th}<3$) calculations based on an $N_g$ adaptive gyroaverage.

\begin{figure}
\begin{center}
\includegraphics[width=3in]{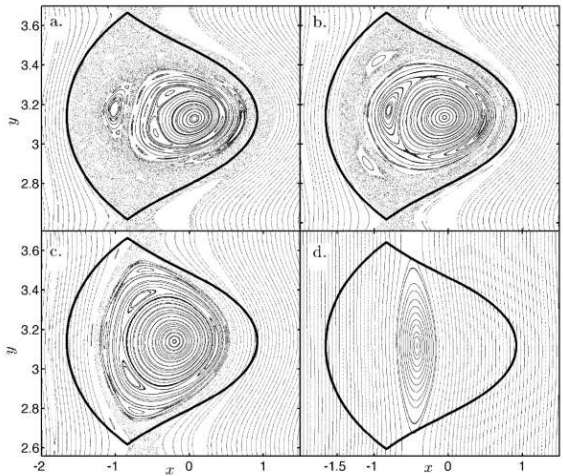}
\end{center}
\caption{Dependence of phase space topology and stochasticity on Larmor radius for the Hamiltonian model in
Eq.~(\ref{eq:stat-6}). The panels show Poincar\'e  maps for a ensemble of particles with
gyroradius distribution of the form $H=\delta(k_{\perp}\rho-k_{\perp}\rho_{th})$ with (a) $k_{\perp}\rho_
{th}=0$, (b) $k_{\perp}\rho_{th}=1.2$, (c) $k_{\perp}\rho_{th}=2.0$ and (d) $k_{\perp}\rho_{th}=3.0$.
The bold, solid curve indicates the unperturbed separatrix for $k_{\perp}\rho_{th}=0$.}
\label{fig:poincare4}
\end{figure}

The Poincar\'e plots in Fig.~\ref{fig:poincare4} show the dependence of the degree of stochasticity on the 
value of $k_{\perp}\rho$. Figure~\ref{fig:poincare4}(a) corresponds to $k_{\perp}\rho=0$. The degree of 
stochasticity is relatively large and, consistent with the results reported in Refs.~\cite{dCN98,dCN00}, the 
stochastic layer is strongly asymmetric. In particular, the region of stochasticity left of the unperturbed 
separatrix (shown with the bold line) is very small. As will be discussed below, this asymmetry manifests 
directly in the skewness of the tail of the test particle propagator, which decays strongly for $\delta  y >0$ 
due to the very low probability of having sticky-flight particles jumping in the $y>0$ direction. 
It may be interesting to compare $\rho_{th}$ to the thickness of the lower branch of the stochastic region, 
$\Delta_s$.  For example, when $k_{\perp}\rho_{th} = \{1.2,2\}$, $\rho_{th}/\Delta_s = \{0.44,1.8\}$.  This 
trend is mainly due to the rapid shrinkage of the stochastic layer as a function of $\rho_{th}$.  When $k_
{\perp}\rho_{th} = 3$, the value of $\Delta_s$ is very difficult to determine because the stochastic layer 
has almost completely disappeared.   

\begin{figure}
\begin{center}
\includegraphics[width=3in]{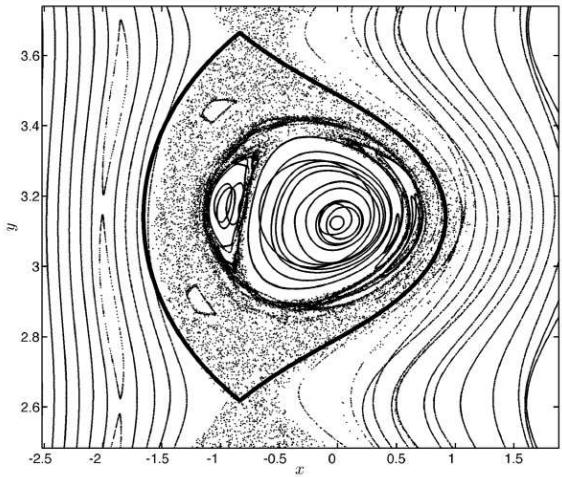}
\end{center}
\caption{Poincar\'e plot for multiple gyroradii values from the Maxwellian distribution with 
$k_{\perp}\rho_{th}=0.6$. Crossings of curves indicate the presence of multiple 
Hamiltonian systems, one for each value of $\rho$.  }
\label{fig:poincarenu06}
\end{figure}

In the FLR calculations the  test particles have a Maxwellian distribution of Larmor radii characterized by 
a mean value, $\rho_{th}$. Thus, depending
on its specific value of $\rho$, each particle ``sees" a different Hamiltonian, which in general will be 
stochastic to a lesser degree as $\rho$ increases.  Figures~\ref{fig:poincare4}(b)-(d) illustrate this with 
Poincar\'e plots corresponding to (b) $k_{\perp}\rho =1.2$, (c) $k_{\perp}\rho =2.0$
and (d) $k_{\perp}\rho =3.0$. Each one of these Poincar\'e sections was computed by assigning the
{\it same} value of $k_{\perp}\rho$, to all the initial conditions.  It is observed that the value of
$k_{\perp}\rho$ has a direct non-trivial influence on the degree of stochasticity. In general, a
Poincar\'e plot corresponding to an ensemble of particles with a Maxwellian distribution of
gyroradii will be a mixture of $k_{\perp}\rho$ Poincar\'e plots, as seen in Fig.~\ref{fig:poincarenu06}.  
The crossings of curves in the Poincar\'e plots indicates the presence of multiple Hamiltonian systems 
indexed by values of $k_{\perp}\rho$.  

\begin{figure}
\begin{center}
\includegraphics[width=3in]{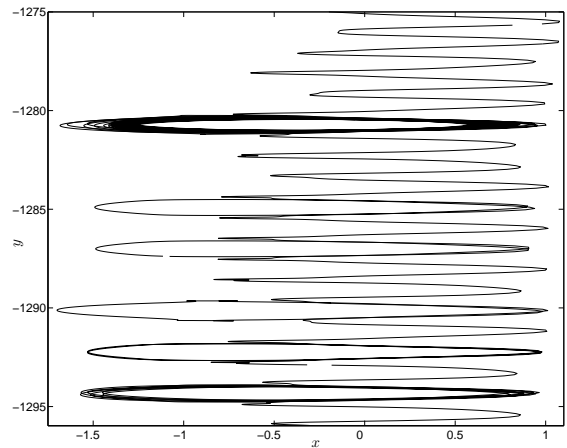}
\end{center}
\caption{Typical sticky-flight trajectory in the Hamiltonian transport model. This particle alternates in a
seemingly unpredictable way between being trapped in ${\bf E} \times {\bf B}$ eddies and being
transported following the zonal shear flow. Other types of orbits, not shown, correspond to trapped
orbits that never leave the original eddy, or passing orbits that move following the zonal flows without
being trapped.}
\label{fig:onetraj}
\end{figure}

To compute the Lagrangian diagnostics of non-diffusive transport, we considered groups of particles
located in the vicinity of a hyperbolic equilibrium point of the Hamiltonian.
The resulting trajectories can be divided into three categories:
(a) passing trajectories that  follow the zonal flow and never enter an ${\bf E} \times {\bf B}$ eddy
(vortex), (b) stagnant trajectories which never leave an eddy and (c) sticky-flight trajectories
which, as shown in Fig.~\ref{fig:onetraj}, alternate between the eddies and the zonal flow. Since the 
statistics of the
passing and the stagnant trajectories are trivial, these particles will be ignored during the data analysis.

Several techniques for isolating sticky-flight trajectories can be devised.  Our trajectory filter works by
examining all trajectories during their entire history, and discarding those that never encircle
a vortex (passing) and those that do not move more than one vortex width from ther original positions
(stagnant).  We have also tested a filter in Fourier-velocity space that discards horizontal velocity time
series without a broadband spectrum.  Depending on the threshold for defining ``broadband," the 
Fourier filter gives practically the same results as the trajectory filter.
Analysis of sticky-flights in more realistic velocity fields would be served better by a
Fourier-velocity filter.  The proper threshold for defining a ``broadband" spectrum can be found from
asymptotic considerations.

\begin{figure}
\begin{center}
\includegraphics[width=3in]{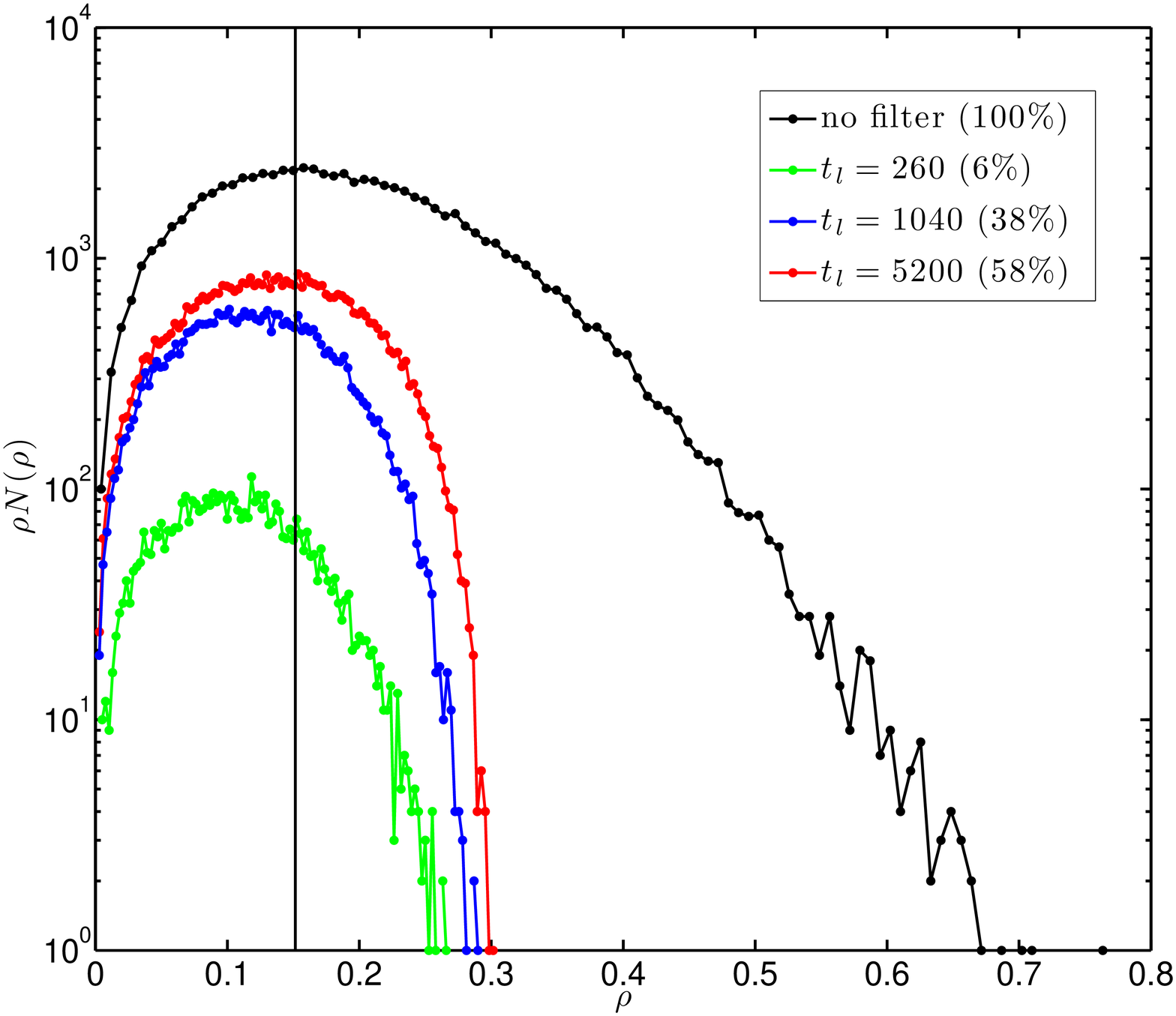}
\end{center}
\caption{(Color online) Gyroradius histogram for $k_{\perp}\rho_{th}=1.2$ with sticky-flight filter applied at various 
times. The uppermost curve shows the unfiltered distribution obtained from the sampling of the $2$-D 
Maxwellian distribution in Eq.~(\ref{eq:maxwell}). The other curves give the distribution at different times 
after the filter (which keeps only the sticky-flight orbits) has been applied.  The vertical line marks the 
maximum of the unfiltered distribution.}
\label{fig:gyroraddist_Kbox}
\end{figure}

Figure~\ref{fig:gyroraddist_Kbox} shows the effect of the trajectory filter on the histogram of Larmor radii. 
In the computation of the histogram we show the number of particles, $N$, multiplied by the appropriate 
metric factor $\rho$. The solid line denotes the histogram considering all the particles in the ensemble, 
{\it i.e.} without the filter. As expected, this histogram corresponds to a sampling of the Mawellian 
distribution in Eq.~(\ref{eq:maxwell}). It is observed that the filter tends to remove particles with large $
\rho$, and, as expected, the number of particles removed decreases with $t_l$, the time of filter 
application. Since $t_l=5200$ appears to give an asymptotic value for the number of sticky-flights, it is 
used as the filtering time for the following diagnostics.  When scaling values are reported for $t<5200$, 
the filter is still applied uniformly at $t=5200$. The first column in Table~\ref{tab:nudeptanh} gives $\Pi_s
$, the percentage of sticky-flights, for each tested value of $k_{\perp}\rho_{th}$ when the filter is applied 
at $t_l=5200$.  

\begin{table}[htdp]

\centering

\caption{Measures of sticky trajectories and non-diffusive transport for the $v_y= \tanh(x)$ model with 
initial positions in a box centered on a hyperbolic fixed point.  The percentage of sticky trajectories at 
$t=5200$, $\Pi_s$, is shown, along with the mean and variance time power law exponents, $\chi$ and $
\gamma$ respectively, at early and late time.  ``Early" refers to a fit for $104<t<1040$ and ``late" refers to 
$4700<t<5200$.  Accuracy for these fits is similar to that observed in Fig.~\ref{fig:meanvarslope}, and 
equal to $\pm0.1$.} 

\begin{tabular}{ c  c  c  c  c  c  c  c }
\hline \hline
$k_{\perp}\rho_{th}$ & $\Pi_s$  & $\chi_{early}$ & $\chi_{late}$ & $\gamma_{early}$ & $\gamma_{late}$ 
& $\zeta_{t=1040}$ \\
\hline  
0.0      & 96 & 1.1 & 1.0 & 1.9 & 1.6  & 2.0\\  
0.001 & 96 & 1.1 & 1.0 & 1.9 & 1.6  & 2.0 \\
0.01   & 96 & 1.1 & 1.0 & 1.9 & 1.6  & 2.0 \\
0.1     & 98 & 1.1 & 1.1 & 2.0 & 1.8  & 2.2 \\
0.2     & 97 & 1.1 & 1.1 & 2.0 & 1.8  &  2.3 \\
0.4     & 96 &  1.1 & 1.1 & 1.9 & 1.9  & 2.3 \\
0.6     & 92 & 1.1 & 1.0 & 1.9 & 1.9  & 2.7 \\ 
0.8     & 83 & 1.1 & 1.1 & 1.9 & 1.9  & 2.7 \\
1.2     & 58 & 0.9 & 1.0 & 1.8 & 1.8  & 2.9 \\
1.6     & 36 & 0.8 & 1.0 & 1.8 & 1.8  & 2.9 \\
3.0     & 11 & 0.9 & 0.9 & 1.8 & 1.6 & 3.1 \\
\hline \hline
\end{tabular}

\label{tab:nudeptanh}
\end{table}

\subsection{Super-diffusive scaling}

Before presenting the chaotic transport results, it is instructive to go back to the simple parallel flow in 
Eq.~\ref{eq:oned} to explore the role of FLR effects on particle dispersion in the context of an integrable 
flow for a ensemble of particles initially distributed according to $P=\delta(x-x_0) \delta(y-y_0)$.
If all the particles have the {\it same} Larmor radius, i.e. if $H(\rho)=\delta(\rho-\rho_{th})$, then as Eq.~
\ref{eq:onedpropdrift} in Appendix \ref{sec:staticdetails} shows, $P$ maintains its delta function shape 
and simply drifts with the effective velocity $J_0(k \rho) U_0$, which in the limit of zero Larmor radius 
corresponds to the parallel flow velocity. In this case, FLR effects are irrelevant since they simply rescale 
the velocity. However, when the particles have different Larmor radii, as in the Maxwellian case of Eq.~
\ref{eq:propagstatic}, the effective velocity of each particle will be different and the initial delta function 
will spread in space as is evident in the particle propagators shown in Fig.~\ref{fig:benchmarks}. In this 
case,  the first and second moments are $M=V_{eff} t$ and $\sigma^2 = A t^2$,
where $V_{eff}$ and A are functions of $k_{\perp} \rho_{th}$ given in Appendix \ref{sec:staticdetails}. 
The key issue to observe is that the variance does not exhibit diffusive scaling, and that a distribution of 
Larmor radii gives rise to a ballistic spreading of the particles.

\begin{figure}
\begin{center}
\includegraphics[width=3in]{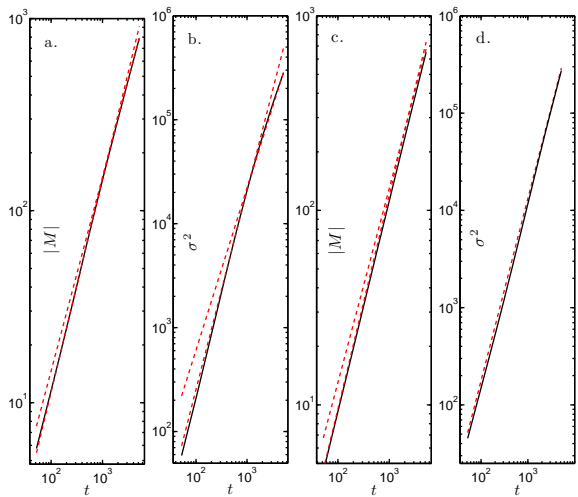}
\end{center}
\caption{(Color online) Time evolution of statistical moments of particle displacements.
Panels (a) and (b) correspond to $k_{\perp}\rho_{th}=0$ and panels (c) and (d) correspond to
$k_{\perp}\rho_{th}=0.6$. Plots (a) and (c) give the absolute value of the first moment $M$, and
plots (b) and (d) show the second moment.  The dashed lines in panels (a) and
(c) have slopes corresponding to $\chi=1.1(0.9)$ and $\chi=1.0$ indicative of
normal advection scaling, {\it i.e.} $|M|\sim t^{\chi}$ with $\chi \approx 1$. The variance shows
super-diffusive scaling {\it i.e.} $\sigma^2\sim t^{\gamma}$ with $\gamma \neq 1$.
However, in the $k_{\perp}\rho_{th}=0$ case, a sharp transition is observed in the anomalous
diffusion exponent. The dashed lines in panels (b) have slopes corresponding to $\gamma=1.9$ and
$\gamma=1.6$. The dashed line in panel (d) has a slope corresponding $\gamma=1.9$ indicating a
uniform scaling of the variance for $k_{\perp}\rho_{th}=0.6$.}
\label{fig:meanvarslope}
\end{figure}

For transport in the nonintegrable flow with the zonal flow and drift waves, 
Fig.~\ref{fig:meanvarslope} shows the mean, $M(t)$, and variance, $\sigma^2(t)$,
for $k_{\perp}\rho_{th}=0$ and $k_{\perp}\rho_{th}=0.6$. A summary of the values of the scaling
exponents $\chi$ and $\gamma$ for all the values of $k_{\perp}\rho_{th}$ studied is presented in
Table~\ref{tab:nudeptanh}.  To a good approximation, the mean exhibits linear scaling, {\it i.e.}
$\chi\approx 1$ in Eq.~(\ref{eq:an_sca}), indicative of regular advection, for all values of $k_{\perp}\rho_
{th}$.  The variance consistently shows clear evidence of super-diffusive transport,
{\it i.e.} $\gamma > 1$ in Eq.~(\ref{eq:an_sca}). In the zero Larmor radius case, two scaling regimes are 
observed. Up to $t \approx
10^3$, which corresponds to the simulations in Ref.~\cite{dCN98},
the power law fitting in Fig.~\ref{fig:meanvarslope}(b) indicates an almost ballistic scaling with
$\gamma=1.9$. However, at a later time there is a transition to $\gamma=1.6$.
As Table~\ref{tab:nudeptanh} shows, FLR effects seem to eliminate the distinction between early and
late regimes.  In particular, according to Fig.~\ref{fig:meanvarslope}(d) where
$k_{\perp}\rho_{th}=0.6$, the scaling $\gamma=1.6$ holds throughout the integration time.
As a general trend, it is observed that the exponent $\gamma$ decreases with increasing
$k_{\perp}\rho_{th}$ beyond $0.1$.  Statistics for sticky-flights become poor for $k_{\perp}\rho_{th}=3$ 
because the degree of stochasticity [see Fig.~\ref{fig:poincare4}(d)] becomes small.

\subsection{Asymmetric, non-Gaussian PDF of particle displacements}
\label{sec:PDF}

\begin{figure}
\begin{center}
\includegraphics[width=3in]{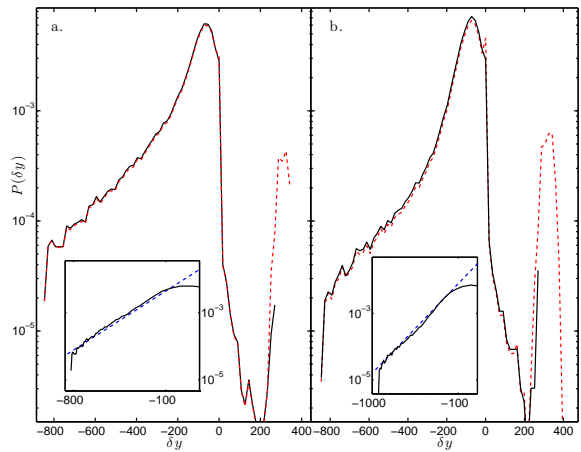}
\end{center}
\caption{(Color online) Probability distribution function of particle displacements at intermediate times,
$t=1040$.  Panel (a) corresponds to  $k_{\perp}\rho_{th} = 0$ and panel (b) corresponds to
$k_{\perp}\rho_{th} = 1.2$. The insets in both figures show evidence of algebraic decaying tails, $P \sim
\delta y^{-\zeta}$ with  $\zeta = 1.95$ for $k_{\perp}\rho_{th} = 0$ and
$\zeta = 2.9$ for $k_{\perp}\rho_{th} = 1.2$.  In both plots, the solid line denotes the PDF of
sticky-flights ({\it i.e.}, excluding the passing and trapped orbits), and the dashed line denotes the PDF
computed using all the orbits.}
\label{fig:dispPDFt1box}
\end{figure}

Motivated by the presence of two different scaling regimens in the variance, we study the PDF of particle
displacements at intermediate and large times. Figure~\ref{fig:dispPDFt1box} shows the PDFs at
intermediate times, with \ref{fig:dispPDFt1box}(a) corresponding to $k_{\perp}\rho_{th}=0$ and \ref
{fig:dispPDFt1box}(b) corresponding to
$k_{\perp}\rho_{th}=1.2$. The solid lines denote the PDFs of the filtered data, ({\it i.e.} including only
sticky-flight orbits) and the dashed line denotes the PDFs of the unfiltered data.
The spikes for large $\delta y$ in the unfiltered distributions result from the contribution of passing
orbits that the filter effectively removes.  The filtered PDFs are clearly
non-Gaussian with strong skewness in the negative $\delta y$ direction.
The strong left-right asymmetry of the PDFs results from the
asymmetry of the stochastic layer.  

In particular, as the Poincar\'e plots in Fig.~\ref{fig:poincare4} show,
the stochastic layer is thicker on the right side of the vortex. This asymmetry depends on the value
of the perturbation frequency $\omega$ in Eq.~(\ref{eq:stat-6}). In fact, as discussed in
Ref.~\cite{dCN98}, the relative thickness of the stochastic layers, and therefore the symmetry of tracer
transport, can be controlled by changing
$\omega$.  As the insets in Fig.~7 show, both PDFs decay algebraically as in
Eq.~(\ref{eq:zeta}). However, a strong dependence of the decay exponent on the value of the Larmor
radius is observed. For
$k_{\perp}\rho_{th}=0$, $\zeta \approx 1.95$,  and for
$k_{\perp}\rho_{th}=1.2$, $\zeta \approx 2.9$.  As Table~\ref{tab:nudeptanh} indicates, the value of the 
decay exponent $\zeta$ increases monotonically with $k_{\perp}\rho_{th}$.

\begin{figure}
\begin{center}
\includegraphics[width=3in]{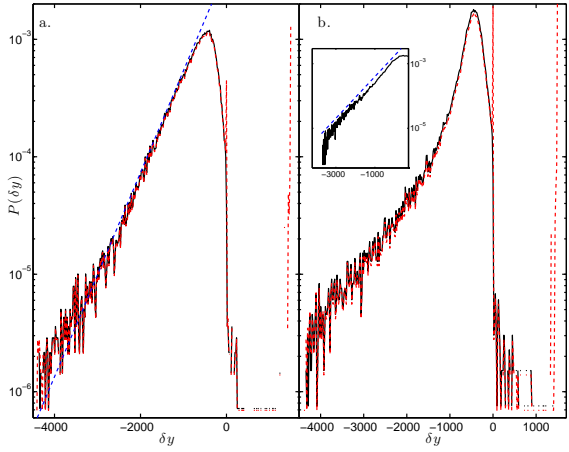}
\end{center}
\caption{(Color online) Probability distribution function of particle displacements at large times,
$t=5200$.  Panel (a) corresponds to  $k_{\perp}\rho_{th} = 0$ and panel (b) corresponds to
$k_{\perp}\rho_{th} = 1.2$. In case (a) the PDF decays exponentially, $P \sim e^{-\lambda \delta y}$
with $\lambda \sim 0.002$. On the other hand, for  $k_{\perp}\rho_{th} = 1.2$,
the inset shows evidence of algebraic decay, $P \sim \delta y^{-\zeta}$
with  $\zeta = 2.9$. In both plots, the solid line denotes the PDF of
sticky-flights ({\it i.e.}, excluding the passing and trapped orbits), and the dashed line denotes the PDF
computed using all the orbits.}
\label{fig:dispPDFt2box}
\end{figure}

The particle displacement PDFs at longer times are shown in Fig.~\ref{fig:dispPDFt2box}. As before, the 
solid lines denote the filtered distribution and the dashed lines the unfiltered distribution. A critical 
dependence on the Larmor radius is observed.  For $k_{\perp}\rho_{th}=0$ the PDF transitions to an 
exponential decaying distribution, whereas for $k_{\perp}\rho_{th}=0.6$ the PDF maintains its algebraic 
decay with the same exponent as the one observed at short times, $\zeta \approx 2.9$.
The robustness of the algebraic decay in the finite Larmor radius case might be attributed to the
persistence of large particle displacements which, due to the presence of the strong zonal flows,  are
enhanced by the gyroaverage.  One should note that a L\'evy process requires $\zeta<3$, which means 
that the increase of $k_{\perp}\rho_{th}$ moves the process away from the L\'evy type.  

The transition from algebraic to exponential decay in the zero Larmor
radius case is likely due to the presence of truncated L\'evy flights. Exact L\'evy flights produce long 
particle displacements that result in slowly decaying algebraic tails at all times. However, non-ideal 
effects such as particle decorrelation might preclude the existence of arbitrarily long displacements, 
resulting in a faster than algebraic decay of the tails at long times.  See, for example, Refs.~\cite
{Mantegna94,Koponen95,Cartea07} for more details on truncated L\'evy processes.  One obvious 
reason for a truncated L\'evy process in the present system is the finite velocity requirement, which 
precludes the existence of infinite jumps.

\subsection{L\'evy flights and algebraic trapping PDFs }

\begin{figure}
\begin{center}
\includegraphics[width=3in]{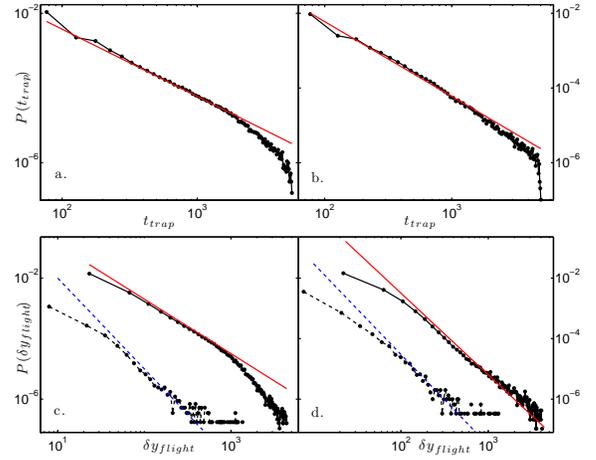}
\end{center}
\caption{(Color online) Probability distribution functions of particle trapping events and particle flight events for
$k_{\perp}\rho_{th} = 0$ and $k_{\perp}\rho_{th} = 1.2$.
The trapping PDFs are shown in (a) and (b), and the flight PDFs
are shown in (c) and (d).  Panels (a) and (c) correspond to $k_{\perp}\rho_{th} = 0$, and
panels (b) and (d) correspond to $k_{\perp}\rho_{th} = 1.2$.
The solid straight lines in (a) and (c) indicate that the trapping PDFs show algebraic decay, $P \sim
t_{trap}^{-\nu}$, with $\nu\approx 1.8$ for  $k_{\perp}\rho_{th} = 0$, and
$\nu\approx 2.0$ for  $k_{\perp}\rho_{th} = 1.2$.
The negative flights PDF shown fit with solid lines also exhibit algebraic decay of the form
$P \sim t_{flight}^{-\mu}$ with $\mu \approx 1.8$ for the case $k_{\perp}\rho_{th} = 0$, and
$\mu \approx 2.7$ for the case $k_{\perp}\rho_{th} = 1.2$. The PDFs of positive flights, shown fit with
dashed lines, show a faster exponential-type decay with $\mu \approx 3.0$ in both cases.}
\label{fig:traptflightx}
\end{figure}

Figure~\ref{fig:traptflightx} shows the trapping time and flight length PDFs for
$k_{\perp}\rho_{th}=0$ in  (a) and (c),  and for $k_{\perp}\rho_{th}=1.2$ in (b) and (d).
In both cases, the trapping PDF clearly decays algebraically as in Eq.~(\ref{eq:trapflight}), with  $\nu 
=1.8$ for $k_{\perp}\rho_{th}=0$, and $\nu =2.0$ for $k_{\perp}\rho_{th}=1.2$.
Figures \ref{fig:traptflightx}(c) and \ref{fig:traptflightx}(d) show the PDFs of flight lengths. Note that, 
because transport in this case is asymmetric, there are actually two flight PDFs, one corresponding to 
positive flights (with dashed fit line) and another corresponding to negative flights (solid fit line).
The PDF of negative flights decays as a power law with
$\mu =1.8$ for $k_{\perp}\rho_{th}=0$, and $\mu =2.7$ for $k_{\perp}\rho_{th}=1.2$.
Since $\mu <3$ in both cases, these flights correspond to L\'evy flights.
However, the decay of the curve for positive flights is much steeper with $\mu \gtrsim 3$ regardless of
the value of $k_{\perp}\rho_{th}$, which implies that positive displacements are not L\'evy flights.
The tails of the trapping and flight PDFs transition to exponential decay at $\delta y_{flight}
\approx -1000$ and $t_{trapt} \approx 2000$.  As discussed before, this transition is indicative of
the possible presence of truncated L\'evy flights.

\subsection{Algebraic decay of Lagrangian velocity autocorrelation function}

\begin{figure}
\begin{center}
\includegraphics[width=3in]{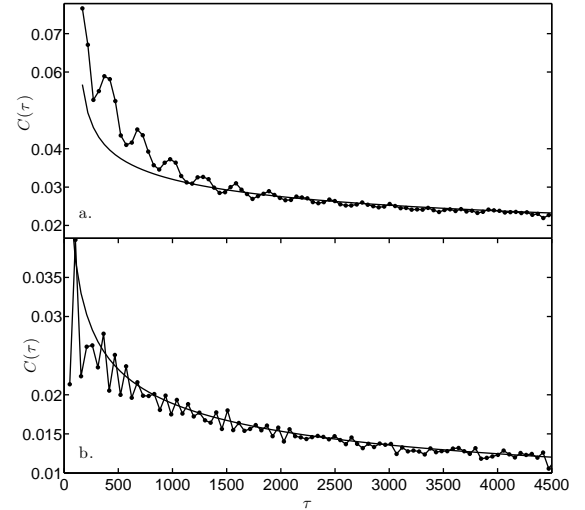}
\end{center}
\caption{Lagrangian velocity autocorrelation function for sticky-flight trajectories. Panel (a) corresponds
to $k_{\perp}\rho_{th}=0$ and panel (b) corresponds
to $k_{\perp}\rho_{th}=1.2$. The curves with dots are the numerical results, and the solid line
curves are algebraic fits of the form ${\cal C} \sim \tau^{-\kappa}$ with $\kappa =0.2$ in (a) and
$\kappa =0.3$ in (b).}
\label{fig:velocityC}
\end{figure}

Figure~\ref{fig:velocityC} shows the Lagrangian velocity autocorrelation function for the sticky-flights 
with $k_{\perp}\rho_{th}=0$ in Fig.~\ref{fig:velocityC}(a) and with $k_{\perp}\rho_{th}=1.2$ in 
Fig.~\ref{fig:velocityC}(b).  Both curves follow algebraic decay of the form $C(\tau) \sim \tau^{-\kappa}$. 
When 
$k_{\perp}\rho_{th}=0$, $\kappa =0.2$ and when $k_{\perp}\rho_{th}=1.2$,  $\kappa =0.3$. Both values 
are consistent with the Green-Kubo relation  between the decay of the velocity correlation and the 
scaling of the variance according to which $\kappa=2-\gamma$. The 
frequency of small scale oscillations observed in the correlation seems to increase when $k_{\perp}
\rho_{th}$ changes from $0\to1.2$.

\section{Self-similar anomalous scaling and fractional diffusion modeling}
\label{sec:modeling}

An important goal of transport modeling is to construct  effective transport equations that describe the
``macroscopic" coarse grained dynamics when given information at the ``microscopic" kinetic level. 
When the microscopic dynamics involves Gaussian, Markovian stochastic processes ({\it e.g.}, a 
Brownian random walk) the macroscopic dynamics can be modeled using diffusion type equations. 
This is the basic idea behind the use of diffusive models to describe collisional transport. 
However, in recent years it has been shown that the standard diffusion picture can fail when non-
Gaussian and/or non-Markovian statistics are present. 

In particular, experimental, numerical and analytical transport studies in fluids and plasmas ({\it e.g.}
Refs.~\cite
{BGZ97,VenkAntOtt97,Weeks98,DMZ98,dCN98,Kovalyov00,dCN04,Benzekri06,Prants06,Hauff07b} 
and references therein) have shown that underlying stochastic processes governing particle transport in 
flows with coherent structures, like zonal flows and eddies, typically involve anomalously large particle 
displacements induced by the zonal flows and/or anomalous particle trapping in eddies. The presence 
of large particle displacements can invalidate the Gaussianity of displacement distributions.  Particle
trapping can introduce waiting time effects that invalidate the Markovian assumption because of
memory effects.  The statistics of particle transport discussed in the previous section shows clear
evidence of these type of phenomena.  This section presents an effective macroscopic model that 
describes quantitatively the spatio-temporal evolution of the PDF of particle displacements.

\begin{figure}
\begin{center}
\includegraphics[width=3in]{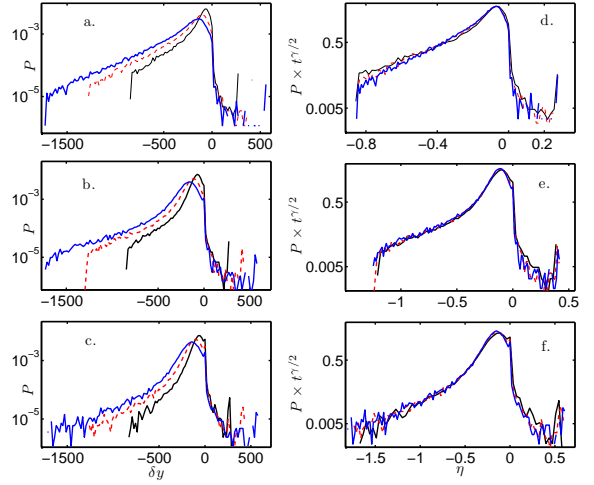}
\end{center}
\caption{(Color online) Self-similar scaling of probability distribution function of particle displacements (PDF).
The curves denote the PDFs at $t=1040$,
$t=1560$, and $t=2080$, with later times showing more spreading in the PDF. Panels (a), (b) and (c) 
show the PDFs corresponding to $k_{\perp}\rho_{th}=0$, $k_{\perp}\rho_{th}=0.6$ and $k_{\perp}\rho_
{th}=1.2$, respectively. Panels (d), (e) and (f) show the collapse of the corresponding PDFs when plotted 
as functions of the similarity variable $\eta=\delta y/t^{\gamma/2}$ and rescaled with the factor $t^
{\gamma/2}$.}
\label{fig:selfsimx3box}
\end{figure}

An important piece of information needed for constructing an effective transport model is shown in Fig.~
\ref{fig:selfsimx3box}. Figures~\ref{fig:selfsimx3box}(a)-(c) show the temporal evolution of the PDF of 
particle displacements for different values of $k_{\perp}\rho_{th}$. As discussed before, the PDF 
develops a strong ``fat" tail to the left and, by conservation of probability, the peak of the
distribution goes down. Figures~\ref{fig:selfsimx3box}(d)-(f) show the same data plotted using rescaled 
variables as in Eq.~(\ref{eq:self_sim_sca}). In the horizontal axis, $\eta=\delta y/ t^{\gamma/2}$, and in 
the vertical axis, $P$ has been multiplied by the
factor $t^{\gamma/2}$, where $\gamma$ is the anomalous diffusion exponent in Eq.~(\ref{eq:an_sca}).
From this it follows that the PDF at a time $\lambda t$ is related to the PDF at time $t$ by the scaling 
transformation $P(\delta y, \lambda t)= \lambda^{-\gamma/2} P(y/\lambda^{\gamma/2}, t)$.
The fact that, for the problem of interest here,   $\gamma \ne 1$, rules out the possibility of
constructing a transport model based on the diffusion equation with an effective diffusivity because the
solution of the diffusion equation scales as $P=t^{-1/2} L(\delta y/t^{1/2})$.

A natural way to built transport models that display self-similar anomalous scaling is to use fractional
diffusion equations of the from
\bq
\label{eq:fde}
_0^CD_t^{\beta}P= \chi_f[l\ _{-\infty}D_y^{\alpha} + r\ _yD_{\infty}^{\alpha}]P \, ,
\eq
where $l=-\sec(\alpha\pi/2) (1-\theta) /2$, and
 $r=-\sec(\alpha\pi/2) (1+\theta) /2$.
The operators $_{-\infty}D_y^{\alpha}$ and $\ _yD_{\infty}^{\alpha}$ are called the left and right
fractional derivatives. These non-local operators are a natural generalization of the regular differential 
operator, $\partial_y^n$, of integer order $n$.  For example, Fourier transforms of the fractional operator,
${\cal F}[f]={\hat f}=\int e^{i k y} f dy$, satisfy
\bq
\label{eq:fourier_fdo}
{\cal F} \left[ _{-\infty}D_y^{\alpha} P \right] = \left(-i k \right)^\alpha {\hat P}  \, ,
\qquad
{\cal F} \left[ _{y}D_{\infty}^{\alpha} P \right] = \left(i k \right)^\alpha {\hat P}  \, ,
\eq
for  non-integer values of $\alpha$.
In a similar way, the operator on the left hand side of Eq.~(\ref{eq:fde}) is a natural extension of the
regular time derivative, $\partial_t f$,  in the sense that its Laplace transform,
${\cal L}[f]= {\tilde f}=\int e^{-s t} f dt$, satisfies
\bq
\label{eq:laplace_caputo}
{\cal L}\left[ _0^CD_t^{\beta} P \right] = s^\beta {\tilde P} - s^{\beta-1} P(t=0) \, ,
\eq
for $0<\beta<1$.
As expected, Eq.~(\ref{eq:fde}) reduces to the  standard diffusion equation
when $\alpha = 2$ and $\beta = 1$.
 Further formal details on fractional derivatives, including their representation in the $y$ and $t$ 
domains
in terms of non-local operators can be found in Refs.~\cite{Podlubny99,Samko93}. For a discussion on 
the use
of these operators to model non-diffusive transport in plasmas, see for example
Refs.~\cite{dCN04,dCN06} and references therein.

To explore the self-similarity properties of the fractional diffusion model we
use Eqs.~(\ref{eq:fourier_fdo})-(\ref{eq:laplace_caputo}) and write the  Fourier-Laplace 
transform, ${\hat {\tilde{\cal G}}}$, of the Green's function,
${\cal G}$, of Eq.~(\ref{eq:fde}) as
\bq
\label{eq:green_fde}
{\hat {\tilde{\cal G}}} = \frac{s^{\beta-1}}{s^\beta - \Lambda}\, , \qquad
\Lambda= \chi_f \left[ l \left(-i k \right)^\alpha + r \left(i k \right)^\alpha \right]  \, ,
\eq
where $\alpha \neq 1$ and ${\cal G}(y,t=0)=\delta(y)$.  It follows directly from Eq.~(\ref{eq:green_fde}) 
that
${\hat {\tilde{\cal G}}}(k,s/\lambda )=\lambda {\hat {\tilde{\cal G}}}(\lambda^{\beta/\alpha} k ,s)$
which in $y$-$t$ space implies the self-similar scaling
${\cal G}(y,\lambda t)=\lambda^{-\beta/\alpha} {\cal G}(\lambda^{-\beta/\alpha} y ,t)$
of the fractional diffusion propagator Eq.~(\ref{eq:fde}). Therefore, the fractional equation will exhibit the 
same self-similar
scaling as the numerically obtained PDF provided the fractional orders of the spatial and temporal
derivatives satisfy
\bq
\label{eq:ratio}
\gamma=2 \beta/\alpha \, .
\eq

According to Table~1, to a good approximation, $\gamma \approx 2$ in the intermediate asymptotic 
regime. Based on this observation, and following Eq.~(\ref{eq:ratio}), we will assume $\alpha=\beta$ in 
the fractional diffusion model.  This special case, known as neutral fractional diffusion, has a Green's 
function that can fortunately be expressed in
closed form using elementary functions, as shown in Ref.~\cite{Mainardi01}:
\begin{eqnarray}
\label{eq:nfde}
{\cal G(\eta;\alpha,\hat{\theta})}= \frac{1}{\pi}\frac{\sin \left[ \pi(\alpha-\hat{\theta})/2 \right] \eta^{\alpha -1}}
{1+2\eta^{\alpha}\cos \left[ \pi(\alpha-\hat{\theta})/2 \right ] + \eta^{2\alpha}}\, , \nonumber\\ {\rm for}
\qquad \eta >0 \, ,
\end{eqnarray}
where $\eta=\delta y/t^{\gamma/2}$ is the similarity variable and $| \hat{\theta}| \leq {\rm min} \{
\alpha, 2-\alpha \}$. The solution for $\eta <0$ is  obtained using the relation
${\cal G}(-\eta; \alpha, \hat{\theta}) ={\cal G}(\eta; \alpha, -\hat{\theta})$. The parameter $\hat{
\theta}$ is related to the asymmetry parameter $\theta$ introduced before in the definition of the
weighting factors $l$ and $r$ according to
$\theta=\tan (\pi\hat{\theta}/2)/\tan (\pi \alpha /2)$.
Given the Green's function, the solution of the fractional diffusion equation for an initial condition
$P_0(\delta y)=P(\delta y,0)$ is
\bq
\label{eq:ic}
P(\delta y,t) = \int_{-\infty}^{\infty} P_0(\delta y')\, {\cal G}(\delta y- \delta y',t)  d \delta y' \, .
\eq

\begin{figure}
\begin{center}
\includegraphics[width=3in]{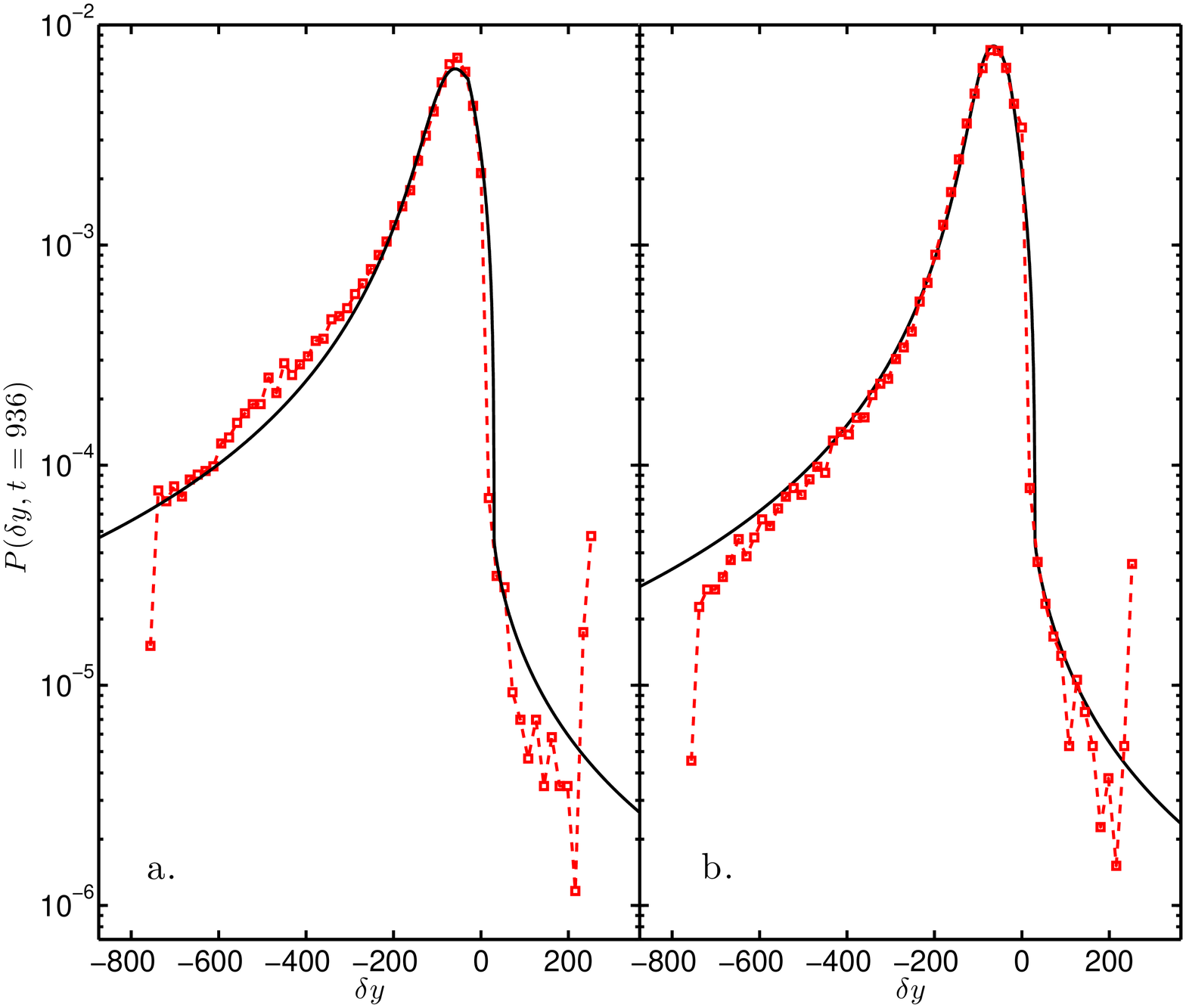}
\end{center}
\caption{(Color online) Comparison between the numerically determined PDF of particle displacements (square 
markers) and the solution of the effective fractional diffusion model in Eq.~(\ref{eq:nfde})(solid lines) with 
a localized initial condition.
In panel (a), which  corresponds to  $k_{\perp}\rho_{th}=0$, the fractional diffusion parameters are
$\alpha =0.8$, $\hat{\theta} = 0.79$, $A= 60$ and $\chi_f = 0.15$. For the case 
$k_{\perp}\rho_{th}=0.6$, shown in panel (b), $\alpha=\beta=.85$, $\hat{\theta}=0.84$, $A=60$ and
$\chi=0.12$.}
\label{fig:epsilonfitbox}
\end{figure}

For the initial condition  we assume a localized distribution of the form
$P_0=1/A$ for $|\delta y|<A/2$ and $P_0=0$ elsewhere (see Ref.~\cite{dCN04}).  The
use of this initial condition is necessary to account for the presence of transients in the evolution of the
PDF not reproduced by the fractional diffusion equation, which describes the intermediate
time regime.  Figure~\ref{fig:epsilonfitbox} shows the comparison of the solution of the fractional diffusion 
equation in
Eq.~(\ref{eq:fde}) according to  Eqs.~(\ref{eq:ic}) and (\ref{eq:nfde}) and the numerically obtained PDF 
obtained
from the histograms of particle displacements at $t=936$ for $k_{\perp}\rho_{th}=0$ in Fig.~\ref
{fig:epsilonfitbox}(a) and
$k_{\perp}\rho_{th}=0.6$ in Fig.~\ref{fig:epsilonfitbox}(b). For the fractional diffusion model parameters 
we used $\alpha=\beta=0.80$ and $\hat{\theta}=0.79$ in the $k_{\perp}\rho_{th}=0$ case, and
$\alpha=\beta=0.85$ and $\hat{\theta}=0.84$ in the $k_{\perp}\rho_{th}=0.6$ case. In both cases,
we used $A=60$, which is small compared to the maximum range of the PDF, $\delta y \sim -800$.

\section{Summary and conclusions}
\label{sec:conclusion}

In this paper we presented a numerical study of FLR effects on non-diffusive transport of
test particles in a flow dominated by a strong zonal shear flow and large scale ${\bf E} \times {\bf B}$
eddies.  We modeled the flow using a  Hamiltonian dynamical system consisting of a 
linear superposition of a strong zonal shear flow and eigenmodes of the HM equation. For 
the parameter values considered, the Hamiltonian causes chaotic transport.  Test particles
alternate stochastically between being trapped in the vortices and being transported by the
zonal flow. To expose the non-diffusive properties of the system we used Lagrangian statistical
diagnostics including: (i) the scaling in time of statistical moments; (ii) the
PDFs of particle displacements, (iii) trapping events and (iv) flight events; and (v) the decay of the 
Lagrangian velocity autocorrelation function.

Finite Larmor radius effects were incorporated in the particle calculations by substituting the value of the
${\bf E} \times {\bf B}$ velocity at the location of the guiding center by its value averaged over a ring of 
radius $\rho$, where $\rho$ is the Larmor radius. The ring average was computed using a discrete 
approximation. The numerical method was benchmarked using an analytical solution for a parallel zonal 
flow with no waves.  We found that for $k_\perp \rho <3$ an $8$-point average gives accurate results, 
but higher order approximations must be used for for $k_\perp\rho >3$. Contrary to previous works 
where all the particles were assumed to have the same value of 
$\rho$, here we considered a more realistic Maxwellian distribution of Larmor radii. Poincar\'e plots 
revealed that the Larmor radius has a direct nontrivial effect on the topology of the flow and the degree 
of chaos of test  particles. In particular, it was observed that the amount of chaos, measured by the width 
of the stochastic region, is significantly reduced as $k_{\perp}\rho_{th}$ increases from $0$ to $3$. A 
distribution of Larmor radii can also have a direct effect on the dispersion of particles. In particular, we 
have shown that, even in the case of a completely integrable flow, particles exhibit ballistic spreading, $
\sigma^2 \sim t^2$, when they have different Larmor radii.

For the Lagrangian statistics we limited attention to sticky-flight orbits and ignored trapped and
passing orbits. The rationale for this filter is that the trivial dynamics of passing and trapped particles give
rise to outliers that artificially bias the statistics.  The first moment, to a good 
approximation, has normal advective scaling, {\it i.e.} $M \sim t^\chi$, with $\chi \approx 1$,  and the 
second moment has super-diffusive scaling, {\it i.e.} $\sigma^2 \sim t^\gamma$, with $\gamma >1$. For
$k_\perp \rho=0$, a sharp transition was observed in the scaling exponent, from $\gamma=
1.9$ at intermediate times to $\gamma =1.6$ at larger times. Similar transitions in the value of $\gamma$ 
have been also found in other systems including temporally irregular channel flows \cite{VenkAntOtt97},  
time dependent, three dimensional flows \cite{Fogleman01}, and two-dimensional vortex flows \cite
{Kovalyov00}. For specific experimental instances, early time behavior will be more important than late 
time behavior if the domain crossing time is small enough.  We have found that FLR effects seem to 
eliminate the distinction between early and late time. For the range of $k_\perp \rho_{th}$ considered, $
\gamma \approx 1.8 \pm 0.1$.  We refer to this regime as super-diffusive ballistic transport since the 
variance approaches ballistic scaling ($\gamma = 2$) but the PDF of displacements retains a super-
diffusive appearance.  Complementary results were obtained in Ref.~\cite{Annibaldi00} for nonlinear 
HM simulations.  

We also observed that the Lagrangian velocity autocorrelation function decays algebraically, ${\cal C} 
\sim \tau^{-\zeta}$ where, in reasonable agreement with the Green-Kubo scaling, $\zeta=2-\gamma$.
The trapping and flight distributions show algebraic decay. The trapping time exponent, $\nu$, remains 
the same when $k_{\perp}\rho_{th}$ changes.  The PDFs of negative flights qualify as truncated L\'evy 
distributions but positive flights are definitively not L\'evy.  The negative flight exponent for $k_\perp \rho_
{th} =1.2$ is larger than expected in the context of a CTRW.
    
At intermediate times, consistent with Refs.~\cite{dCN98,dCN00}, the PDF of particle displacements in
the zero Larmor radius case is an asymmetric non-Gaussian distribution with an algebraic decaying 
leftward tail.  However, for larger times, the tail of the PDF transitions from algebraic to
exponential decay. This algebraic-exponential transition in the PDF is likely to be related to 
the presence of truncated L\'evy flights, which, as discussed in Ref.~\cite{Cartea07}, might result from
particle decorrelation or the finite size of possible displacements. The robustness of the algebraic decay 
in the finite  Larmor radius case might be attributed to the persistence of large particle displacements 
which, due to the presence of the strong zonal flows,  are enhanced by the gyroaverage. We have also 
shown that the PDF of particle displacements has self-similar scaling behavior for $0 \leq k_{\perp}\rho_
{th}\leq 3$ and $k_{\perp}\rho_{th}\neq 0$. Most importantly, we have shown that these distributions  
correspond  to solutions of the neutral ($\alpha=\beta$) asymmetric fractional diffusion equation.

Future work will apply the ideas and tools developed here to turbulent flows to more realistic plasma 
turbulence models.  In particular, we will examine self-consistent particle transport parallel to a density 
gradient in a gyrokinetic particle-in-cell simulation.  Transport properties of tracers and self-consistent 
particles should be compared. 

\begin{acknowledgments}

Thanks to T.M. Antonsen, Jr., S. Brunner, P. Ricci, M. Barnes and I. Broemstrup for helpful discussions. 
This work is supported by the Fannie and John Hertz Foundation. Additional support comes from the 
Oak Ridge National Laboratory, managed by UT-Battelle, LLC, for the U.S. Department of Energy under 
Contract DE-AC05-00OR22725 and from the DOE Center for Multiscale Plasma Dynamics, Grant DE-
FC02-04ER54784.

\end{acknowledgments}


\begin{appendix}

\section{Gyro-averaged particle propagator in a parallel flow}
\label{sec:staticdetails}

The gyroaverage equations of motion for test particles in the
parallel zonal flow of
Eq.~\ref{eq:oned} are
\begin{eqnarray}
\frac{d x}{d t}= 0\, , \qquad
\frac{d y}{d t}= - \phi_{0} k_{\perp} \left <  \sin (k_{\perp} x)\right >_{\theta}= \nonumber\\
-\phi_{0} k_{\perp} J_{0}(k_{\perp} \rho) \sin (k_{\perp}  x) \, .
\end{eqnarray}
A straightforward integration assuming an intial condition
$(x_0,y_{0})$ gives
\bq
x=x_{0} \, ,\qquad
y=y_{0} -   U_{0} J_{0}(k_{\perp} \rho) \, t\, ,
\eq
where $U_{0}=\phi_{0} k_{\perp} \sin (k_{\perp}  x_{0})$.
From here it follows that the two-dimensional propagator is
\bq
\label{eq:onedpropdrift}
{\cal P}({\bf r}, t | {\bf r'}, t'; \rho) = \delta (x-x')\,\delta \left[ y -y' + J_0(k_{\perp} \rho)  U_0 t\right ]
 \, .
\eq
Integrating over $x$ and assuming a Maxwellian
distribution of gyroradii gives the one-dimensional propagator in
$y$,
\begin{eqnarray}
\label{eq:propagstatic}
P(y,t|y',t';\rho) = \nonumber\\ \frac{2}{\rho_{th}^2} \int_0^\infty  \delta \left[ y
-y' + J_0(k_{\perp} \rho)  U_0 t
\right ] \rho \, e^{-\rho^2/\rho^2_{th}} \,\, d \rho  \, .
\end{eqnarray}
Integrating over $\rho$ using basic properties of the delta function
gives  Eq.~\ref{eq:exact}.
From Eq.~(\ref{eq:propagstatic}) it follows that the $n$-th moment of the gyrocenter
displacement $\delta y= y-y'$ scales like $t^n$ according to
\bq
\langle (\delta y)^n\rangle= (U_0 t)^n \int_0^\infty J_0^n(k_{\perp}\rho)\,
H(\rho) d \rho \, .
\eq
where $H(\rho)$ is the gyroradii distribution function.
For $n=1$ and $n=2$ we recover the moments in Sec.~\ref{sec:results}(A) with
\begin{eqnarray}
V_{eff}=U_{0} e^{-k_{\perp}^{2}\rho^{2}_{th}/4}\, \nonumber\\
A=U_{0}^{2} e^{-k_{\perp}^{2}\rho^{2}_{th}/2}\left[ I_{0}\left(
k_{\perp}^{2} \rho_{th}^{2}/2\right)-1\right]
\end{eqnarray}
in the case when $H$ is Maxwellian,
where $I_{0}$ is the modified Bessel function of zero-order. It is
interesting to note that $A$ has a maximum for $k_{\perp}\rho_{th}
\approx 2.5$.

\end{appendix}

\bibliographystyle{prsty}
\bibliography{kbgmaster}

\end{document}